\documentstyle[12pt,epsfig,rotating]{article}
\def\bsls35{\baselineskip 0.35in}

\def\bd{\begin{document}} \def\ed{\end{document}}
\def\bmp{\begin{minipage}} \def\emp{\end{minipage}}
\def\bcc{\begin{center}} \def\ecc{\end{center}}     \def\npg{\newpage}
\def\beq{\begin{equation}} \def\eeq{\end{equation}} \def\hph{\hphantom}
\def\be{\begin{equation}} \def\ee{\end{equation}} \def\r#1{$^{[#1]}$}
\def\n{\noindent} \def\ni{\noindent} \def\pa{\parindent}
\def\hs{\hskip} \def\vs{\vskip} \def\hf{\hfill} \def\ej{\vfill\eject}
\def\cl{\centerline} \def\ob{\obeylines}  \def\ls{\leftskip}
\def\underbar#1{$\setbox0=\hbox{#1} \dp0=1.5pt \mathsurround=0pt
   \underline{\box0}$}   \def\ub{\underbar}    \def\ul{\underline}
\def\f{\left} \def\g{\right} \def\e{{\rm e}} \def\o{\over}
\def\vf{\varphi} \def\pl{\partial} \def\cov{{\rm cov
}} \def\ch{{\rm ch}}
\def\la{\langle} \def\ra{\rangle} \def\EE{e$^+$e$^-$}
\def\bitz{\begin{itemize}} \def\eitz{\end{itemize}}
\def\btbl{\begin{tabular}} \def\etbl{\end{tabular}}
\def\btbb{\begin{tabbing}} \def\etbb{\end{tabbing}}
\def\beqar{\begin{eqnarray}} \def\eeqar{\end{eqnarray}}
\def\\{\hfill\break} \def\dit{\item{-}} \def\i{\item}
\def\bbb{\begin{thebibliography}{9} \itemsep=-1mm}
\def\ebb{\end{thebibliography}} \def\bb{\bibitem}
\def\bpic{\begin{picture}(260,240)} \def\epic{\end{picture}}
\def\akgt{\noindent{\bf Acknowledgements}}
\def\fgn{\noindent{\bf\large\bf Figure captions}}
\def\fsz{\footnotesize}
\def\ifmath#1{\relax\ifmmode #1\else $#1$\fi}%
\def\rmt{\ifmath{{\mathrm{t}}}} \def\rmcut{\ifmath{{\mathrm{cut}}}}
\newcommand{\QCD}{{\sc qcd}} \newcommand{\NFM}{{\sc nfm}}
\newcommand{\BNL}{{\sc bnl}} \newcommand{\RHIC}{{\sc rhic}}
\newcommand{\CERN}{{\sc cern}} \newcommand{\LHC}{{\sc lhc}}
\newcommand{\ALICE}{{\sc alice}}
\def\pt{{p_{\rmt}}} \def\vf{\varphi} \def\yct{y_{\rmcut}} \def\kt{k_{\rmt}}
\def\levy{L$\acute{\rm e}$vy} \def\renyi{R$\acute{\rm e}$nyi}

\begin{document}

%

%

\cl{\Large Comparison of the Geometrical Characters Inside }
\vskip0.2cm \cl{\Large Quark- and Gluon-jet Produced by Different
Flavor Quarks}

\vskip0.4cm \cl{ \  Chen Gang$^{(1,2)}$ \quad Wei Hui-ling$^{(1)}$}

\vskip0.4cm \cl{\small $^{(1)}$ School of Mathematics and Physics,
China University of Geosciences, Wuhan, China, 430074}

\cl{\small $^{(2)}$ Key Laboratory of Quark \& Lepton Physics
(CCNU), Ministry of Education,China,430079}

\footnotetext[0]{chengang1@cug.edu.cn}

\begin{abstract}
The characters of the angular distributions of quark jets and gluon
jets with different flavors are carefully studied after introducing
the cone angle of jets. The quark jets and gluon jets are identified
from the 3-jet events which are produced by Monte Carlo simulation
Jetset7.4 in $e^+e^-$ collisions at $\sqrt s$=91.2GeV. It turns out
that the ranges of angular distributions of gluon jets are obviously
wider than that of quark jets at the same energies. The average cone
angles of gluon jets are much larger than that of quark jets. As the
multiplicity or the transverse momentum increases, the cone-angle
distribution without momentum weight of both the quark jet and gluon
jet all increases, {\it i.e} the positive linear correlation are
present, but the cone-angle distribution with momentum weight
decreases at first, then increases when $n>4$ or $p_t>2$GeV. The
characters of cone angular distributions of gluon jets produced by
quarks with different flavors are the same, while there are obvious
differences for that of the quark jets with different flavors.
\end{abstract}


{\bf PACS \ \ numbers: 13.87FH,\ \ 13.66.BC}

\section{Introduction}

According to quantum chromodynamics(QCD), the basic elements that
constitute substances are quarks and gluons. Because of  "color
confinement",  we could not find free quarks and gluons. However,
through analyzing the hadronization production from quarks and
gluons, the characters of quarks and gluons can be obtain
indirectly.

In 1975, 2-jet events were found for the first time in $e^+e^-$
collisions\cite{jet2}. After this, in 1979, 3-jet events were
observed in the energy range $17-30$GeV in $e^+e^-$
collisions\cite{jet3}. According to local parton-hadron dualism
(LPHD)\cite{parton}, hadron jets can reflect information of the
decay and hadronization of original partons. Thereby through
studying on jets, we can indirectly get characters of strong
interactions among quarks and gluons.

The viewpoint that flavor quantum number is independent of strong
interaction is the basic properties of QCD, and the only reason for
destroying flavor symmetry is the "mass effect" in heavy quarks
decaying. For the color charge of gluon is larger than that of
quark, gluon jet has the character that it is fatter than quark jet.
The experimental result \cite{experi4}$^{-}$\cite{experi11} obtained
in LEP $e^+e^-$ storage ring (CERN) is quantificationally in
agreement with the theoretically predictions: gluon jet has larger
average charged particle multiplicity, softer fragmentation function
and wider angular distributions than quark jet
\cite{result11}.Recently, the general
characteristics\cite{new13}$^{-}$\cite{new16} of and the dynamical
fluctuations inside\cite{new17}$^{,}$\cite{new18} quark and gluon
jets from three-jet events have been analyzed using Monte Carlo
simulation in $e^+e^-$ collisions.

In this thesis, we study the geometrical characters of quark jets
and gluon jets and compare the differences between quark jets and
gluon jets, starting from the characters of angular distributions;
and analysed and compared carefully the geometrical characters of
jets formed by different flavor quarks and them emitted gluons with
the aim of studying the difference among these kinds of jets, which
the flavor of the original quark is discriminated by changing
parameter of model.

The event samples of final state particles of $e^+e^-$ collisions at
$\sqrt s=91.2$GeV are produced by Monte Carlo Simulation Jetset7.4
generator. The three-jets sub-samples are obtained by Durham
jet-algorithm~\cite{Durham} from the full event samples
\cite{identify}. According to QCD, the three-jets are separately
produced by the hadronization of the original quark and anti-quark
produced in $e^+e^-$ collision and a hard gluon emitted by one of
the original quark (or anti-quark). The original quark (or
anti-quark) which had emitted this hard gluon is called mother
quark. For convenience of comparing characters of quark jets and
gluon jets with the same conditions, in this thesis we studied jets
produced by mother quarks and gluons. Hereinafter the mother quark
jet is just called quark jet for short, and jets formed by quark is
called quark jet with flavor of this quark. For example, jet formed
by b-quark is called b-quark jet. To check whether there is any
character difference among gluons emitted by quarks with different
flavors, we especially distinguished the gluons: a gluon emitted by
one quark with certain flavor is called gluon with this kind of
quark flavor, and correspondingly jet formed through this gluon
hadronization is called gluon jet of this quark kind. Such as, one
hard gluon emitted by a b-quark is called b-gluon; and jet formed
though this b-gluon hadronization is called b-gluon jet.

\begin{figure}[pb]
\centerline{\psfig{file=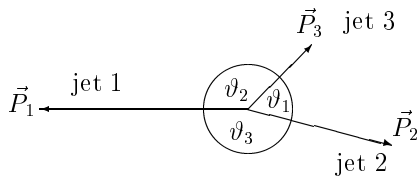,width=5.5cm}} \vspace*{6pt}
\caption{The skeleton sketch of 3-jet event distribution}
\end{figure}

To study the properties of quark jets and gluon jets, we also need
to identify the two quark jets and one gluon jet from the three jets
in a 3-jet event. We chose the angular method \cite{DM}, which can
be simply expressed with fig.1. Where, $P_i(i=1,2,3)$ is the sum of
momentum of all particles in jet-$i$. The angles between each two
jets are defined as
\begin{equation}
\theta_{i}=\arccos\left(\frac{P_{j1}P_{k1}+P_{j2}P_{k2}+P_{j3}P_{k3}}
{P_{j}P_{k}}\right),(i,j,k=1,2,3;i\neq{j},j\neq k, k\neq i).
\end{equation}

Where jet facing the largest angle $\theta_3$ is the gluon jet, jet
facing the smallest angle $\theta_1$ is the jet formed by the
original quark which had not emitted hard gluon, and jet facing the
middle angle $\theta_2$ is the mother quark jet. Considering the
requirement of momentum conservation, thus the three jets should lie
in one plane, we added one condition:
$\theta_1+\theta_2+\theta_3\geq 359^\circ$. To improve the purity of
selected events\cite{YM}, we added condition: $\theta_3-\theta_2\geq
10^\circ$.

The thrust frame(as axis $z$) is commonly used for the
three-dimensional phase space in e$^+$e$^-$
collisions~\cite{thrust}. In order to get the characters of angular
distributions inside quark jets and gluon jets more accurately, the
jet frame is constructed taking the total momentum of jet as the
longitudinal axis (axis $z$)~\cite{yuyan}.

\section{The 2-dimensional angular distributions of particles inside jets}

To describe the angular distribution characters of particles inside
jets, we defined two angular distribution variables $\alpha_1$ and
$\alpha_2$,
\begin{equation}
\begin{array}{c}
\alpha_1=f_1(\theta,\phi)=P_x/P=\sin\theta\cos\phi,\\
\alpha_2=f_2(\theta,\phi)=P_y/P=\sin\theta\sin\phi.
\end{array}
\end{equation}

Where $\theta$ is the angle between the particle momentum direction
and the jet momentum direction, and $\phi$ is the angle between
particle momentum projection in transverse plane and $x$-axis
direction.

\begin{figure}[pt]
\centerline{\psfig{file=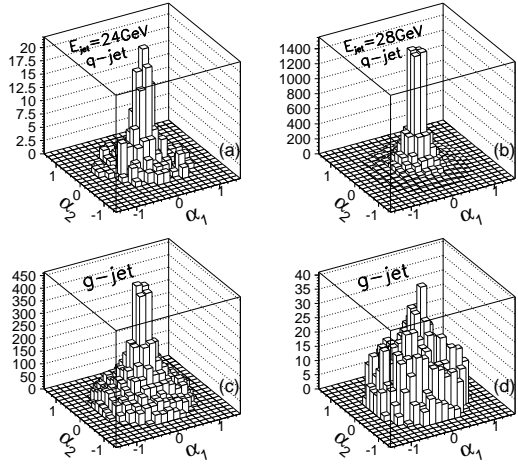,width=8cm}} \vspace*{6pt}
\caption{The 2-dimensional angular distribution column diagram of
particles inside quark jets and gluon jets. The first line is the
quark jet, and the second line is the gluon jet; the first column
the energy is 24GeV, the second column the energy is 28GeV.}
\end{figure}

 In our Monte Carlo simulation, a total number of 5000,000 events of $e^+e^-$ collisions at
91.2 GeV are produced by Jetset 7.4 generator and the numbers of
379,825 3-jet events are selected out using Durham algorithm, then
we separately selected out quark jets and gluon jets with energy at
24GeV, 28GeV from 3-jet events. According to the definition of
angular distribution variables in equation (2), we plot the the
2-dimensional angular distributions of particles in quark jets and
gluon jets with different jet energy, as it is shown in Fig.2. Fig.2
provides us column diagram of 2-dimensional angular distributions of
quark (without distinguishing the original quark flavors) jets and
gluon jets with energy at 24GeV and 28GeV,respectively.

It can be seen from Fig.2 that the angular distribution range of
particles inside gluon jets are obviously wider than that inside
quark jets with the same energy, and the distribution has perfect
symmetry relative to jet axis, namely, the space distribution of
jets present as taper. All these conclusions is consistent with the
predictions of QCD theory.


\section{Definition of cone angle}

In the previous section, we qualitatively discussed the angular
distribution characters of particles inside jets and compared the
angular distribution characters of quark jets and gluon jets. To
quantifacationally study the geometrical characters of jets, jet
cone angle is defined as follows\cite{cone}.

Suppose the number of charged particles of jet-$i$ in one event is
$n_i$, the total momentum of this jet is $P^i$, the momentum of the
$j$-th charged particle in this jet is $P^i_j$, thus the cone angle
of this jet is defined as
\begin{equation}
\langle \theta
\rangle=\frac{1}{n_i}\sum_{j=1}^{n_i}\arccos(\frac{\overrightarrow{P^i}\cdot{\overrightarrow{P_j^i}}}
{|\overrightarrow{P^i}||\overrightarrow{P_j^i}|}).
\end{equation}
Considering status of particles with momentum of different
magnitudes is different in jets, namely, because they have different
effect on the jet, we added momentum weight $\omega_j$ while
calculating cone angles of jets:
\begin{equation}
\omega_j=\frac{|\overrightarrow{P_j^i}|}{\frac{1}{n_i}\sum_{k=1}^{n_i}|\overrightarrow{P_k^i}|}.
\end{equation}
Thus cone angle with momentum weight of jet is defined as
\begin{equation}
\langle \theta
\rangle=\frac{1}{n_i}\sum_{j=1}^{n_i}\omega_j\arccos(\frac{\overrightarrow{P^i}\cdot{\overrightarrow{P_j^i}}}
{|\overrightarrow{P^i}||\overrightarrow{P_j^i}|}).
\end{equation}

\section{The cone angular distributions of quark jets and gluon jets}

In order to study the cone-angle property of jets, four event
samples of final state particles with 5,000,000 events are generated
from Jetset 7.4 generators for e$^+$e$^-$ \ collisions at c.m energy
91.2~GeV. Then the events of number 379825 for 3-jets are obtained
by Durham algorithm from the full event samples, and the single
quark jet and single gluon jet sub-samples are selected from 3-jet
events using the angular rule with energy at 18GeV and 24GeV,
respectively. Using these sub-samples we can analysis and compare
various properties for quark- and gluon- jets.

 The distributions of cone angle defined in Eq's.~(3) and (5) are
shown in Fig.3 for quark- and gluon- jets with energy at 18GeV and
24GeV, respectively. It can be seen from Fig.3~(a), with the energy
of jet at 18GeV, that the cone angle of gluon-jet are distributed
over 1$^\circ$ -- 50$^\circ$ with an average value equal to \
24.6$^\circ$; while the cone angle of quark-jet take values in the
region 1$^\circ$ -- 45$^\circ$ and the average is only \
16.2$^\circ$. When the energy of jet are 24GeV, showing in
Fig.3~(b),the distribution of cone angle for gluon-jet and quark-jet
move slightly to right to that of the cone-angle with energy at
18GeV.

\begin{figure}[pb]
\centerline{\psfig{file=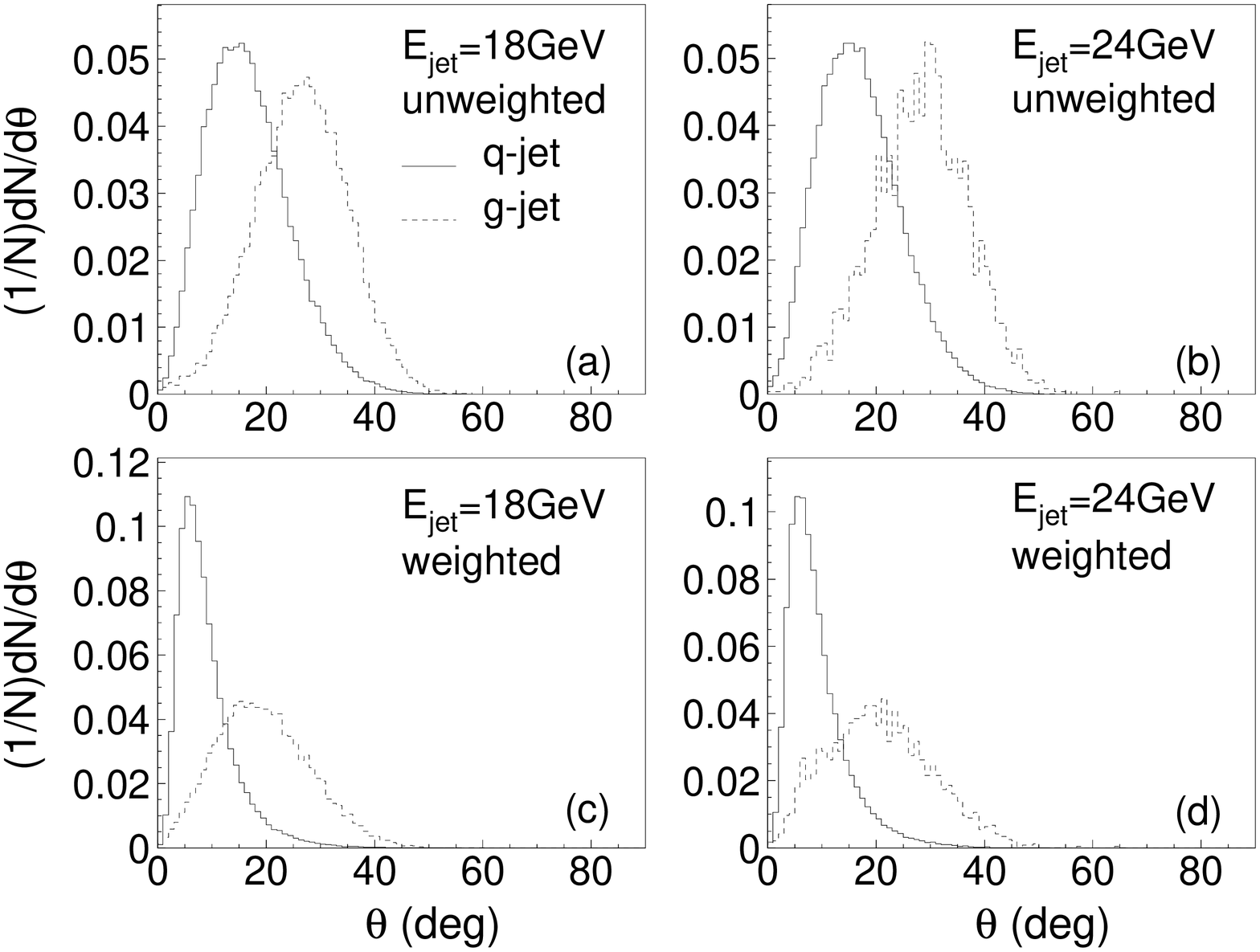,width=9cm}} \vspace*{8pt}
\caption{The cone angular distributions of quark jets and gluon jets
with energy at (a)(c) 18GeV and (b)(d) 24GeV. (a)(b) without
momentum weight, (c)(d)with momentum weight.}
\end{figure}

The distribution of cone-angle of jet with the momentum weight is
shown in Fig.3~(c) and (d). It can be seen through comparing
Fig's.3~(a),((b) and (c),(d) that the distribution of weighted
cone-angle is moved leftward for \ 4.8$^\circ$ -- 6.3$^\circ$ in
comparison to that of the unweighted cone-angle. The reason is
because the momenta of particles nearby the jet axis is generally
bigger than those far from the jet axis. The relation between the
distributions of quark- and gluon- jets are qualitatively the same
for weighted and unweighted cone-angles.

\section{Comparison of correlation characters with average cone
angles}

We use Monte Carlo generator Jetset7.4 to produce 5,0000,000
$e^+e^-$ collision events at an energy $\sqrt s=91.2GeV$. The event
samples are constructed according to the flavors of the original
quark-pair $b\overline b, c\overline c, d\overline d, s\overline s,
u\overline u$, respectively. The 3-jet event sub-samples are
selected out using Durham rules with the cut parameter
$y_{cut}=0.002$, and  the quark-jets(mother quark-jets) and
gluon-jets are selected out from 3-jet events using angular method.

In order to study the correlation of cone angle of jets with other
jet-variables, {\it e.g.} jet multiplicity ---- the number of
charged particles in a jet. we divide the multiplicity region \
$n_{\rm {l}} = 2 \sim$12 into\ 11 bins and separately calculate the
average cone angles of jets in each multiplicity bin.
\begin{equation}
\langle \theta_C \rangle_{l,k}=\frac{1}{n_l}\sum^{n_l}_{m=1}\langle
\theta_C \rangle^{l,m}_{jet-k}, \qquad (k=1,2,3;\,l=2,\cdots,12).
\end{equation}

Where $\langle \theta_C \rangle^{l,m}_{jet_k}$ is the cone angle of
the $m$th jet of kind $k$ with charge multiplicity in the
multiplicity bin $l$th. The $n_l$ is the number of jet in the $l$th
multiplicity bin,$<\theta_{\rm c}>_{l,k}$ is the cone-angle of the
jet of kind $k$ jet in the $l$th bin.The results of average
cone-angle of the gluon jet and quark jet {\it vs}. multiplicity in
each multiplicity bin for Jetset 7.4 Monte Carlo generators are
shown in Fig's~4 (a) and (b), respectively.

\begin{figure}[pb]
\centerline{\psfig{file=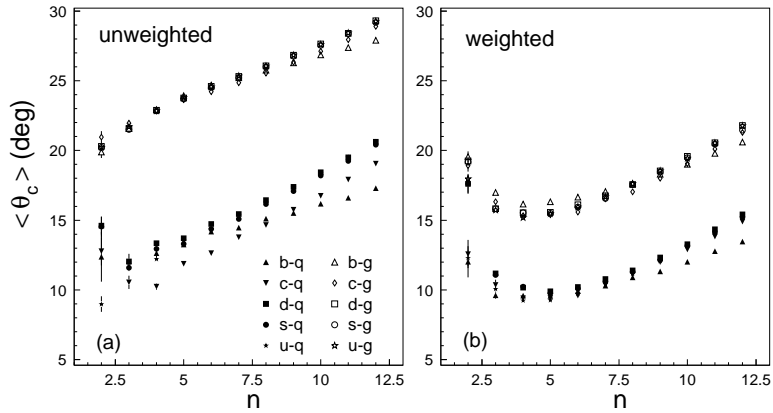,width=10.5cm}} \vspace*{8pt}
\caption{The distributions of average cone angles with different
flavors quark jets and gluon jets as functions of charged
multiplicities. (a) without momentum weight, (b) with momentum
weight}
\end{figure}

It can be seen from the Fig~4 (a) that the average cone-angles of
the gluon jet and quark jet {\it vs}. multiplicity act differently.
The average cone-angle of the gluon jet is bigger than that of the
quark jet, providing further evidence that the gluon jet is "fatter"
than quark jet. The average cone-angle of the quark jet and gluon
jet all increases with multiplicity.

The results of cone-angles with momentum weight and unweight are
some quantificationally difference. The fig~4 (a) shows that the
cone-angle of gluon jet and quark jet without momentum weight
increases linear as the increasing of multiplicity, i.e the
cone-angle of gluon jet and quark jet are positive correlation with
multiplicity. The fig~4 (b) shows that the cone-angle of gluon jet
and quark jet with momentum weight decreases as the increasing of
multiplicity for \ $n < 4~$, and after that the cone-angles of gluon
jet and quark jet increase with the increasing of multiplicity when
$n > 4~$, developing a valley, i.e the cone-angle of gluon jet and
quark jet are negative correlation with multiplicity as $n < 4~$ and
positive correlation as $n > 4~$.
 This appearance existing minimum
values after taking momentum weight into consideration is mainly
aroused by the leading particles effect. When $n=2$, the jet axis
lies between the two particles and the average cone angle is
relatively large; when $n=3$ or $n=4$, as momentum of the leading
particles is larger than the other particles and it closes with the
jet axis, the average cone angles of jet become smaller after
considering momentum weight; when $n>4$, for number of particles
gradually become more, momentum of leading particles become less and
effects to average cone angles become weaker, so the average cone
angle increases with the increasing of charged particle multiplicity
as $n>4$.

 It is remarkable that the
average cone-angle of the gluon jet with different flavors are
equation within range error, which illuminate that gluons jet
produced by hard gluon emitted by quarks with different flavors have
certain same characters; while that of the quark jet with different
flavors are qualitatively conformable and are quantificationally
difference, which illustrates that characters of different flavors
quarks are different. This kind of difference is mainly aroused by
mass difference among different flavors quarks.

Now we turn to the correlation between cone-angle and transverse
momentum. We divide the transverse momentum range $P_t=0 \sim
10$~GeV/c into 12 bins, and calculate the average cone-angle with
different flavors quark jets and gluon jets in each transverse
momentum bin.
 When calculating the average cone angles in each
transverse momentum intervals, we separately made statistics of the
total cone angle of one certain kind of jets in transverse momentum
bin $l$, divided it by the number $n_l$ of this certain kind of jets
in this transverse momentum bin, then we got the average cone angle
of this certain kind of jets in this transverse momentum bin:
\begin{equation}
\langle \theta_C \rangle_{l,k}=\frac{1}{n_l}\sum^{n_l}_{m=1}\langle
\theta_C \rangle^{l,m}_{jet-k}, \qquad (k=1,2,3;\,l=1,2,\cdots,12).
\end{equation}

Where $n_l$ is the number of jet in the $l$th transverse momentum
bin, $\langle \theta_C \rangle^{l,m}_{jet_k}$ is the cone angle of
the $m$-th jet of kind $k$ in the $l$th transverse momentum bin. The
results of average cone angle of the gluon jet and quark jet with
different flavors {\it vs}. transverse momentum in each transverse
momentum bin for Jetset 7.4 Monte Carlo generators are shown in
Fig's~5 (a) and (b), respectively.
\begin{figure}[pt]
\centerline{\psfig{file=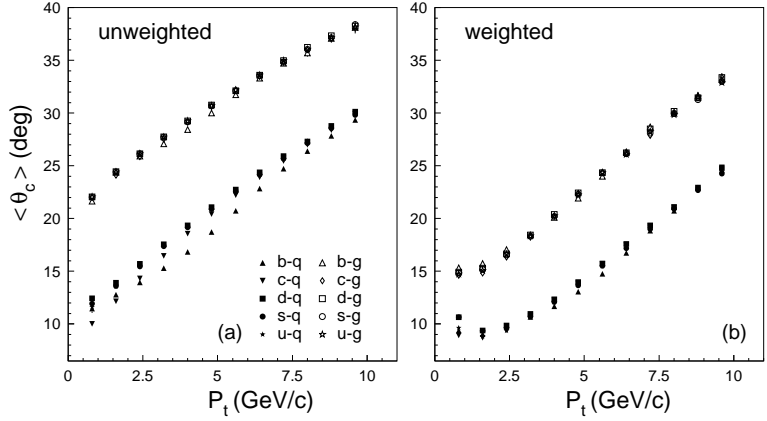,width=10.5cm}} \vspace*{6pt}
\caption{The distributions of average cone angles of quark jets and
gluon jets with different flavors
 as functions of transverse momentum.
(a)\ without momentum weight, (b)\ with momentum weight.}
\end{figure}

It can be seen from the Fig~5 (a) that the average cone-angles of
the gluon jet and quark jet {\it vs}. transverse momentum are
differently, and they increases quickly with the increaes in
transverse momentum for all flavors, showing a strong positive
linearly correlation. The average cone angle of the gluon jet is
bigger than that of the quark jet, providing also evidence that the
gluon jet is "fatter" than quark jet.

The distributions of cone angles with momentum weight and unweight
{\it vs}. transverse momentum are qualitative similar comparing
Fig~4 (a) and (b). When we added momentum weight as Fig~4 (b), their
distributions present properly linear positive correlation only in
higher transverse momentum area; while in lower area their curves
decreased for first one or two points, which may probably be roused
by the leading particle effect as same as section 2.

We also see that the distributions of average cone angle of the
gluon jet with different flavors {\it vs}. transverse momentum are
superposition within range error, which shows that gluons jet
produced by hard gluon emitted by different flavor quarks have
certain same characters; while that of the quark jets with different
flavors are qualitatively conformable and are quantificationally
difference. This kind of difference is mainly aroused by mass
difference among different flavors quarks.

In order to explain the character difference of cone angular
distributions with different flavors quark jets in fig.4 and fig.5,
we still use Monte Carlo simulation Jetset~7.4 to produce data,
divide the transverse momentum range $P_t=0.4 \sim 10$GeV/c into 12
equal bin, and then get distributions of charged multiplicity with
different flavors quark jets as functions of transverse momentum, as
 shown in Fig~6.

From Fig~6, it can be seen that average charged multiplicities of
quark jets with different flavors are different in the same
transverse momentum bin, distributions of average charged
multiplicities arrange to mass magnitude of the original quarks, and
the larger the mass the larger the charged multiplicity is. Suppose
mass of two kinds of quarks are $m_A$ and $m_B$, and the final state
charged hadron multiplicities are $n_A$ and $n_B$, respectively.
Then we have
\begin{figure}[pt]
\centerline{\psfig{file=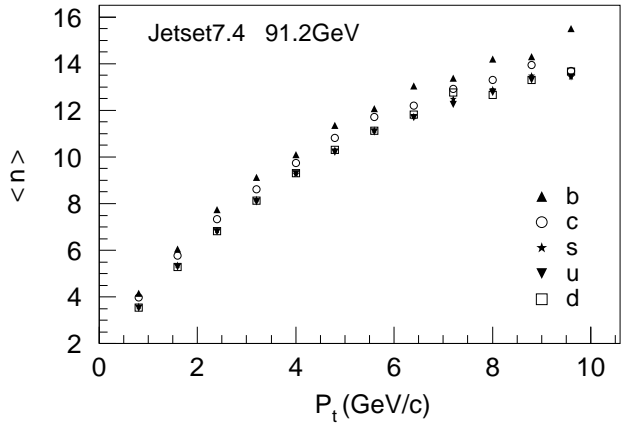,width=9cm}} \vspace*{8pt}
\caption{The distributions of multiplicities with different flavors
quark jets as functions of transverse momentum.}
\end{figure}
\begin{equation}
n_A > n_B,\quad \quad (as~~~ m_A > m_B).
\end{equation}
We introduce a simple model base above result. First we suppose
there are only two kinds of particles in jets: one kind is large
vertical momentum and small transverse momentum, noted as $\Gamma$,
and the transverse momentum of each particle is $P$, the angle
between the particle and the jet axis is $\alpha$; the other is
small vertical momentum and large transverse momentum, noted as
$\Lambda$, and the transverse momentum of each particle is $Q$, the
angle between the particle and the jet axis is $\beta$. Obviously,
\begin{equation}
    \alpha<\beta, \quad \quad  P<Q.
\end{equation}
Arbitrarily choose two different kinds of quarks $A$ and $B$ as the
research objects, their mass separately are $m_A$ and $m_B$, and
$m_A>m_B$. Jet formed by qark-$A$ is noted as $A$-jet and jet formed
by quark-$B$ is noted as $B$-jet. Under the same condition of
transverse momentum $P_t$, suppose the number of $\Gamma$ particle
is $n$, number of $\Lambda$ particle is $m$ in $A$-jet, and suppose
the number of $\Gamma$ particle is $N$ and number of $\Lambda$
particle is $M$ in $B$-jet. Then we have
\begin{equation}
P_t=nP+mQ=NP+MQ.
\end{equation}

The average cone angles of $A$-jet and $B$-jet separately are
\begin{equation}
\langle \theta \rangle_{A-{\rm jet}}=\frac{n\alpha+m\beta}{n+m},
\qquad \langle \theta \rangle_{B-{\rm
jet}}=\frac{N\alpha+M\beta}{N+M}.
\end{equation}
Substituting equation (11) with equations (8)-(10), we obtain
\begin{equation}
\langle \theta \rangle_{A-{\rm jet}} -\langle \theta \rangle_{B-{\rm jet}}=(\alpha-\beta)(nM-Nm)\\
=(\alpha-\beta)(\frac{P(n-N)}{Q})<0.
\end{equation}
Therefor, we get conclusion that the average cone angle of $A$-jet
is smaller than that of $B$-jet with the same transverse momentum,
i.e, the larger the quark mass the smaller the average cone angle
is.

 Suppose rates of number of $\Gamma$ particles and number of $\Lambda$
 particle in one certain kind of jet to the total charged multiplicity of this jet
  is a constant. And note
  rate of number of $\Lambda$ particles in $A$-jet to
  charged multiplicity of this jet is $x$, rate of number of $\Lambda$ particles in
  $B$-jet to charged multiplicity of  $B$-jet is $y$, thus
\begin{equation}
x=\frac{n}{n+m},\quad \quad \quad
y=\frac{N}{N+M}.
\end{equation}
From equation (8)-(10), we have $n>N$ and $m<M$. After substituting
them into equation (13) we get $x>y$, i.e, proportion of number of
particles with large vertical momentum in $A$-jet is higher than
that in $B$-jet. When charged multiplicities $N_{\rm tot}$ for
different flavors quark jets are the same, the average cone angle of
jets can be expressed as
\begin{equation}
\langle \theta
\rangle_{A-jet}^{N_{tot}}=N_{tot}[x\alpha+(1-x)\beta],
\end{equation}
\begin{equation}
\langle \theta
\rangle_{B-jet}^{N_{tot}}=N_{tot}[y\alpha+(1-y)\beta].
\end{equation}
Thus slopes of distribution curves of average cone angles as
functions of charged multiplicities can be got according to the
following equations
\begin{equation}
\frac{\partial{\langle \theta
\rangle_{A-jet}^{N_{tot}}}}{\partial{N_{tot}}}=x\alpha+(1-x)\beta,
\end{equation}
\begin{equation}
\frac{\partial{\langle \theta
\rangle_{B-jet}^{N_{tot}}}}{\partial{N_{tot}}}=y\alpha+(1-y)\beta.
\end{equation}
For $x>y$, thus
$$\frac{\partial{\langle \theta \rangle_{A-jet}^{N_{tot}}}}{\partial{N_{tot}}}<
\frac{\partial{\langle \theta
\rangle_{B-jet}^{N_{tot}}}}{\partial{N_{tot}}}.$$

Namely slope of distribution for $A$-jet is smaller than that for
$B$-jet. For $m_A>m_B$, this means that the larger the mass of quark
for producing jet the smaller the distribution slope of the average
cone angle as function of jet transverse momentum.

For five different flavors quarks $b, c, s, u, d$, mass of them
exist difference, i.e, the mass of quark $b$ (5GeV) is much larger
than mass of other flavors quarks, next is quark $c$ (1.5GeV), quark
$u, d$ has the least mass. In summery: (1) The distributions of
average cone angles {\it vs}. charged multiplicities with different
flavors quark jets are different, shown in fig~4 (a) and (b). For
mass of quark-b is the largest, slope of its distribution is the
least. (2) The distributions of average cone angles {\it vs}.
transverse momentum among different flavors quark jets is some
difference, shown in fig~5 (a) and (b). Since mass of quark-b is the
largest, the average cone angle distributions is the lowest.

\section{ Conclusion}

 In this paper, we produced the data using Monte Carlo simulation Jetset7.4 in $e^+e^-$
collision events with the center of mass energy at 91.2GeV,
 selected out 3-jet events with Durham algorithm and
identifyed quark jets and gluon jets using angular method. We
defined angular variables $\alpha_1$, $\alpha_2$ for qualitatively
describing the angular distribution characters inside jets and
defined cone angles of jets for quantificationally describing the
geometrical characters of particles inside jets.


 The cone angle distribution of gluon jet at the same energy
 is obviously wider than that in quark jets. The average cone
angles of gluon jets, at the same multiplicity or the same
transverse momentum, are much larger than that of quark jets. This
is in agreement with the predictions of QCD theory that gluons are
"fatter" than quarks, providing further evidence that the gluon jet
is "fatter" than quark jet.

The average cone angles of gluon and quark jet without weight
increases with charged multiplicities or transverse momentum, i.e,
they present linear positive correlations. This illustrates that
cone angle can reflect the distribution characters of particles
inside jets, and can also reflect transverse momentum distribution
characters, namely, we can use this geometrical characters to
describe the dynamical characters inside jets.

The distributions of average cone angle of gluon and quark jet with
momentum weight as functions of charged multiplicity or transverse
momentum all have a minimum value. Namely, with the increasing of
charged multiplicity or transverse momentum, the average cone angle
decreases at first, then increases when $n>4$ or $p_t>2$GeV.  The
appearance of minimum value is aroused by the leading particle
effect.

The distribution of average cone angles of the gluon jets with
different flavors
 {\it vs}. multiplicity or transverse momentum is same,
while that of the quark jets has distinctions, which is as a result
of that the mass difference of different flavors quarks raised the
symmetrical broken of strong interaction.


\section*{Acknowledgments}
This work is supported by NSFC under projects 10775056. The authors
thank Professor Yuanfang in Particle Physics Research Institute of
Huazhong Normal University for helpful discussions.


\clearpage


\begin{thebibliography}{}
%

\bibitem{jet2}G. Hanson , G. S. Abrams, A. M. Boyarski et al.,
 {\it Phys. Rev. Lett.} {\bf 35} (1975) 1609.

\bibitem{jet3} R. Brandelik  et al., {\it Phys. Lett. B} {\bf 86} (1979) 243.

\bibitem{parton} Yu. L. Dokshitzer, S. I. Troyan,  In: Proc. of {\it the XIX
Winter School of the LNPI}, Vol. 1, Leningrad (1984) 144 ;
\\Ya. I. Azimov, Yu. L. Dokshitaer, S. I. Khosze,  {\it Z. Phys. C} {\bf 27}
(1985) 65.
\bibitem{experi4}  OPAL Collab. (G. Alexander et al.), {\it Phys. Lett. B} {\bf 265}
(1991) 462.
\bibitem{experi5} OPAL Collab. (P. D. Acton et al.), {\it Z. Phys. C} {\bf 58}
(1993) 387.
\bibitem{experi6} OPAL Collab.(R. Akers et al.), {\it Z. Phys. C} {\bf 68} (1995) 179.
\bibitem{experi7} OPAL Collab.(G. Alexander et al.), {\it Z. Phys. C} {\bf 69}
(1996) 543.
\bibitem{experi8} DELPHI Collab.(P. Abreu et al.), {\it Z. Phys. C} {\bf 70}
(1996) 179.
\bibitem{experi9} ALEPH Collab.(D. Buskulic et al.), {\it Phys. Lett. B} {\bf 346}
(1995) 389.
\bibitem{experi10} ALEPH Collab.(D. Buskulic et al.),{\it CERN-PPE} (1995) 184.
\bibitem{experi11} ALEPH Collab.(D. Buskulic et al.), {\it Phys. Lett. B} {\bf 383}
(1996) 353.
\bibitem{result11}I.G. Knowles et al., Physics at LEP 2, Vol. 2, {\it CERN 96-01}, eds.  Altarelli G and Zwirner F;
Gary J W, Proceedings of the XXV International Symposium on
Multiparticle Dynamics, Star\'{a} Lesna\'{a}, Slovakia, September
(1995) 12-16.

\bibitem{new13} G. Chen and L.-S. Liu, {\it J. Phys. G} {\bf 30} (2004) 1399.
\bibitem{new14} K.S.Zhang et al., {\it High Energy Phys. Nucl. Phys. }{\bf 26} (2002) 1110 (in
Chinese).
\bibitem{new15} G. Chen, L. S. Liu et al., {\it Int. J. M. Phys. E }{\bf 16}(10)
(2007) 3386.
\bibitem{new16} F.G. Tian, G. Chen et al., {\it Int. J. M. Phys. A }{\bf 23}(26)
(2008) 4337.
\bibitem{new17} H. L. Wei, G. Chen et al., {\it Int. J. M. Phys. E} {\bf  17}(8)
(2008) 1467.
\bibitem{new18} G. Chen, M. L. Yu et al., {\it Chinese Science
Bulletin }{\bf 54} (24) (2008) 3808.
\bibitem{new19} G. Chen, K.S. Zhang and L.S. Liu, {\it Science China G} {\bf 46}(3) (2003) 268.
\bibitem{new20} L.S. Liu, G. Chen and Q.H. Fu, {\it Phys. Rev. D} {\bf 63} (2001) 054002.

\bibitem{Durham} Yu. L. Dokshitzer, {\it J. Phys. G} {\bf 17} (1991) 1537.
\bibitem{identify} S. Catani et al., {\it Phys. Lett. B} {\bf 269}
(1991) 432.
\bibitem{DM} M. Derrick et al., {\it Phys. Lett. B }{\bf  165} (1985) 449.
\bibitem{YM} M. L. Yu and L. S. Liu, {\it Chin. Phys. Lett.} {\bf 19}
(2002) 647.

\bibitem{thrust} E. Farhi, {\it Phys. Rev. Lett.} {\bf 39} (1977) 1587.

\bibitem{yuyan} L. Lu, L. J. Yang and L. P. Yang, {\it HEP £¦ NP} {\bf 25}(11) (2001) 1077 (in
chinese).

\bibitem{cone} G. Chen and L. S. Liu,  {\it Chin. Phys. Lett.} {\bf 22},
(2005) 840.
\end{thebibliography}
\end{document}